# Fully Automated Organ Segmentation in Male Pelvic CT Images

Anjali Balagopal[§], Samaneh Kazemifar[§], Dan Nguyen, Mu-Han Lin, Raquibul Hannan, Amir Owrangi, Steve Jiang

Medical Artificial Intelligence and Automation Laboratory, Department of Radiation Oncology, University of Texas Southwestern, Dallas, Texas, USA
Email: Steve.Jiang@utsouthwestern.edu

## ABSTRACT

Accurate segmentation of prostate and surrounding organs at risk is important for prostate cancer radiotherapy treatment planning. We present a fully automated workflow for male pelvic CT image segmentation using deep learning. The architecture consists of a 2D organ volume localization network followed by a 3D segmentation network for volumetric segmentation of prostate, bladder, rectum, and femoral heads. We used a multi-channel 2D U-Net followed by a 3D U-Net with encoding arm modified with aggregated residual networks, known as ResNeXt. The models were trained and tested on a pelvic CT image dataset comprising 136 patients. Test results show that 3D U-Net based segmentation achieves mean (±SD) Dice coefficient values of 90 (±2.0)% ,96 (±3.0)%, 95 (±1.3)%, 95 (±1.5)%, and 84 (±3.7)% for prostate, left femoral head, right femoral head, bladder, and rectum, respectively, using the proposed fully automated segmentation method.

## INTRODUCTION

Accurate segmentation of prostate and surrounding organs at risk (OARs) is important for radiotherapy treatment planning. Manual segmentation by physicians currently used in clinical practice is time consuming, highly depends on the physicians' skill and experience, leading to large inter and intra observer variation [van Mourik et al., 2010; Van Herk, 2004; Weiss and Hess, 2003; Weiss et al., 2003; Rasch et al., 2005; Vorwerk et al., 2009]. Inter-observer variability (standard deviation) for the manually segmented prostate volumes was observed to range from 10% to 18% [C. Fiorino et al.,1998], indicating the high demand for automated segmentation methods in clinical practice. Automated segmentation of prostate and organs from CT images, is challenging for two reasons: 1) The boundary between prostate and background is usually unclear because of low contrast in the CT images. 2) Variation in shape, size, and intensity of prostate, bladder, and rectum among different patients of different ages and at different times are large. Also, pelvic CT images are largely diverse. Sources of this diversity are described as follows: 1) Use of contrast agents in some patients may partially or fully brighten the bladder in CT images. 2) Since the images are often acquired with different fields of view, substantial organ position and volume dimension variation occurs. 3) Sometimes fiducial markers or catheters may be implanted into patients, changing the texture of major pelvic organs. The diversity of pelvic CT images further complicates the segmentation of male pelvic organs from CT images.

[§] Co-first authors.

Segmentation algorithms have been developed to reduce these variabilities and improve accuracy and efficiency [Acosta O et al.,2010, Acosta O et al.,2011, Bueno G et al.,2011,Chai X et al.,2013, Chen S et al,.2009 , Chen T et al,.2009, Cosio FA et al.,2010, Costa MJ et al.,2007, Ghosh P et al.,2008, Rousson M et al.,2005, Tang X et al.,2004, Shi Y et al.,2013, Gao Y et al.,2016, Li W et al.,2011, Ma L et al.,2016, Shao Y et al.,2014, Lay N et al.,2013, Gao Y et al.,2012, Feng Q et al.,2010, Martinez et al.,2014, Ma Ling et al.,2017, Li, D. et al.,2016, Yang, X., et al., 2014]. Most developed methods fall into the category of deformable model-based segmentation, atlas-based segmentation, and learning-based segmentation.

In recent years, the shape model based on prior knowledge and machine learning has been widely used in the automatic segmentation of CT images. These methods use a training dataset as prior knowledge. Even though they methods perform remarkably well, many require training samples from the previous treatment images of the same patient to learn patient-specific information for enhancing the prostate segmentation result. These methods also rely on hand crafted features for accurate segmentation. Despite all the progress, an obvious gap between the automated segmentation results and manual annotations needs to be addressed.

Deep learning and specifically, convolutional neural networks (CNNs) have revolutionized the field of natural image processing. Instead of extracting features in a hand-designed manner, deep learning discovers the informative representations in a self-taught manner from a set of a data with minor preprocessing only [Bengio Y et al.,2009, LeCun Y et al.,2015]. The U-Net [Ronneberger O et al., 2015] is a well-known CNN architecture for medical images segmentation. Their architecture includes so-called skip connections between feature maps in the same depth level of the contracting and expanding path. The concatenation of the feature maps gives U-Net a significant advantage compared to the patch-wise approaches, preserving local features while propagating global features to higher resolution layers.

Recently, deep learning has been used for prostate segmentation in MRI [Zhu, Qikui et al.,2017, Tian, Zhiqiang et al.,2018, Ishioka, Junichiro et al.,2018, Yu, Lequan et al.,2017], achieving good results. MRI provides excellent soft-tissue contrast helping to better delineate organ boundaries. However, because current radiotherapy treatment planning workflow uses CT images for contouring and dose calculation, automated segmentation of prostate and OARs in CT image is highly desirable, despite more challenging than segmentation in MR images.

Ma et al. (2017) used a combination of deep learning and multi atlas fusion for the automatic segmentation of prostate on CT images. A 2D fully convolutional network (FCN) was used for initial segmentation followed by multi-atlas label fusion to refine the segments. The model was trained on a dataset of 92 patients and achieved a Dice similarity coefficient (DSC) value of 86.80% as compared to manual segmentation. In an earlier work, we have used a 2D U-Net for semi-automated segmentation of prostate and OARs in Pelvic CT and achieved a DSC value of 88% compared to manual segmentation (Kazemifar, S, et al. 2018). We present a fully automated workflow for pelvic CT multi organ segmentation that uses a multi-channel 2D U-Net for organ localization, followed by a 3D U-Net with ResNeXt blocks for accurate segmentation of prostate and organs-at-risk in pelvic CT. The significance of this work is two-fold: 1) This is a first fully automated workflow for pelvic CT segmentation using only deep learning for multi organ segmentation, 2) The proposed 3D network provides better segmentation accuracy compared to existing methods, and is easily translatable to clinical practice.

## METHODS

In general, an organ segmentation problem consists of two related tasks: organ volume localization and organ volume delineation. Organ localization determines the target object's whereabouts on the image or its location, whereas organ delineation draws the object's spatial extent and composition. A fully automated network should be capable of completing both tasks with good accuracy.

Only ~15% of the CT slices acquired for a pelvic CT scan for a patient contains Prostate. Identification of these slices are usually done manually even for the automated segmentation methods that exist [Shao Y et al., 2014]. To create a fully automated network, detection and localization of organs should be done automatically without any manual intervention before segmentation. However, Inter-patient variation in shape, size, and organs location hinders organ detection in CT images. Any error in localization accuracy would completely mislead segmentation. To perform organ localization efficiently on Pelvic CT scans, we have included a 2D U-Net in our workflow before segmentation.

### Automation workflow

The proposed framework for the automated detection and segmentation of prostate, bladder, rectum, and femoral heads in pelvic CT scans is shown in Figure 1.

The upper and lower 15% of the image does not contain any useful information or organs. However, because it increases memory consumption it has been cropped out. The cropping margin was determined for the patients scanned at UT Southwestern Medical Center. The value should be determined according to the dataset. This step was used only for increasing the computing speed and could be removed because it does not affect accuracy.

The images are then downsized before being fed into the localization network to speed up the network and to overcome memory issues. The entire CT volume is fed into the localization network to detect organ locations. Organ volumes are then extracted and fed into the final segmentation network.

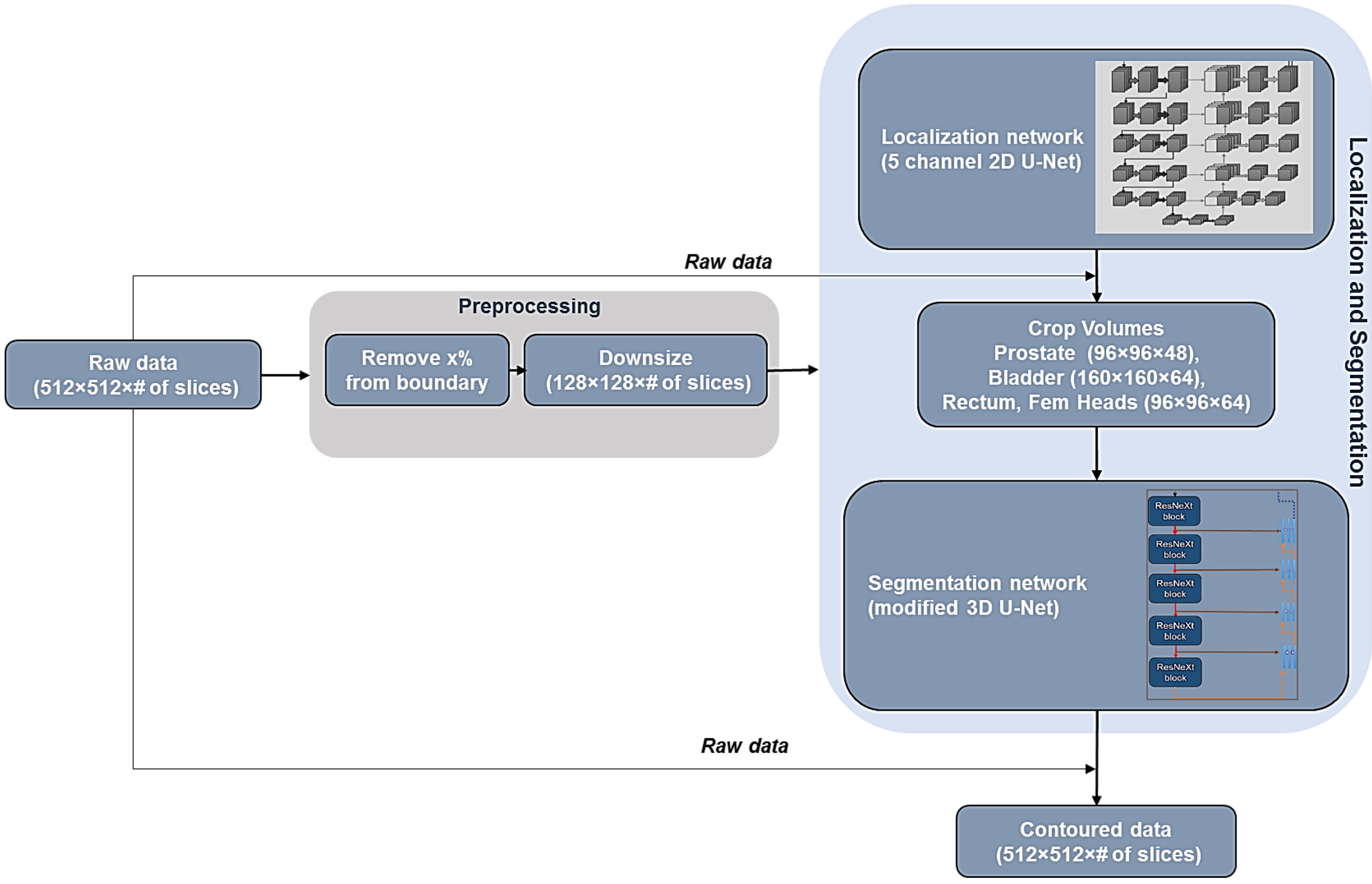

*Figure 1: Automation workflow that requires minimal data preprocessing for organ localization and segmentation*

## Localization network

The localization network used is a 5 channel 2D U-Net architecture (Figure 2) where each channel corresponds to one of the organs to be localized. The CT and binary mask (for supervised training) corresponding to each organ is input separately into each channel. The final predicted output will have 5 channels with each channel corresponding to the predicted feature map for one organ. The network was trained by minimizing a custom weighted Dice loss function, $DSC_w$ (describe in the section on loss function), using the Adam optimizer [D. P. Kingma et al., 2014].

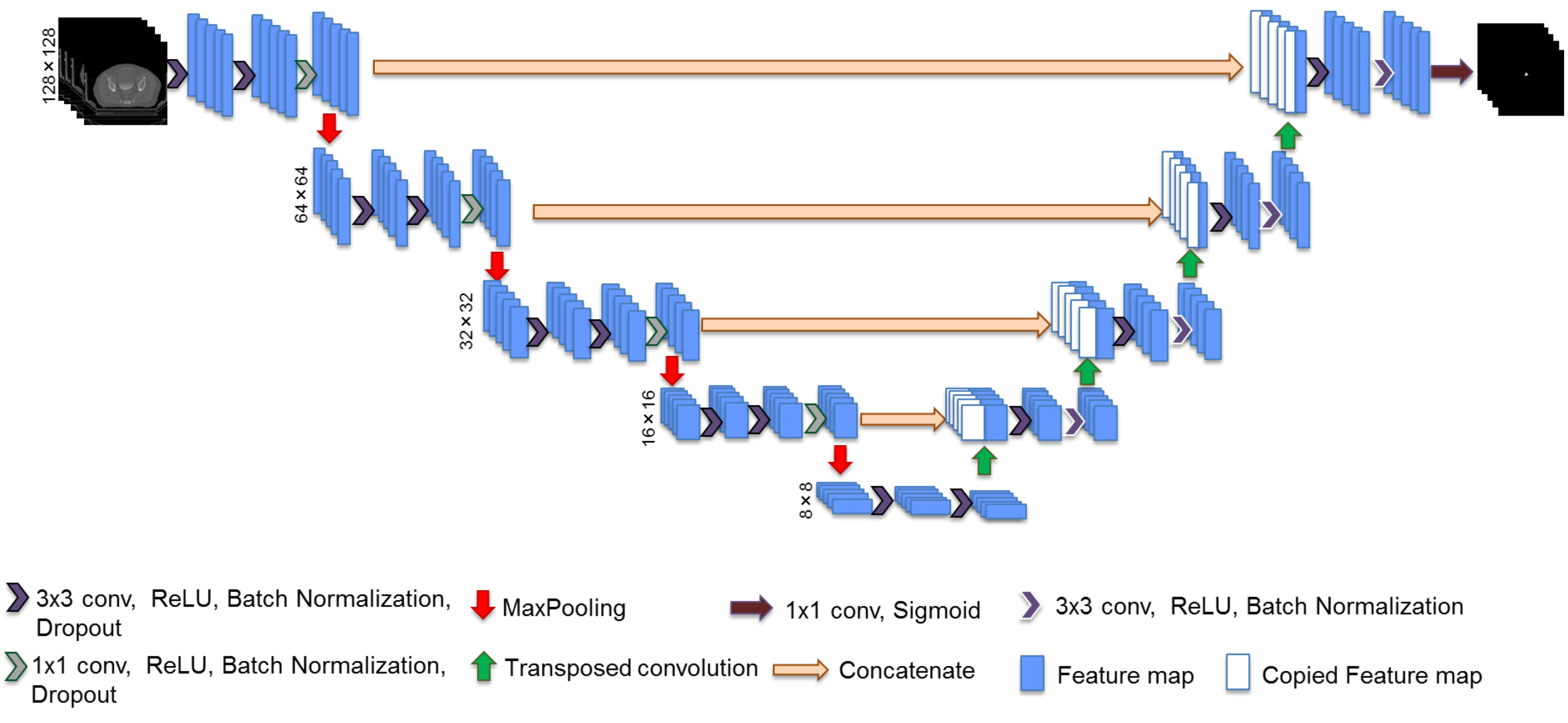

*Figure 2: Localization network: a 5 channel 2D U-Net*

**Segmentation network**

For most organs, because the first and last few slices do not contain sufficient shape information, they would be difficult to segment using 2D networks. Applying 3D kernels would generate discriminative feature maps that contain 3D shape and surface of the related organ rather than generating 2D curves only. A three-dimensional network could comprehend the continuity of sequential slices and also could better segment small parts of the organs compared to 2D networks.

We compared different variations of UNet for an optimal architecture selection. The employed U-Net architecture(Figure 3) is the 3D extension described by Cicek et al. (2016), developed from the U-Net proposed by Ronneberger O et al. (2015) with the encoding arm consisting of ResNeXt blocks adapted from S. Xie et al. (2017) at each layer. Xie et.al. [S. Xie et al., 2017] introduced the concept of "cardinality" – an additional dimension to depth and width- and showed that aggregating residual blocks with the same topology and hyper-parameters is more effective in gaining accuracy than going deeper or wider. The ResNeXt block used in our network consists of $n$ repetitions of two 3D convolutions (filter size, 3×3×3) and is represented in Figure 3. The outputs of these $n$ paths are concatenated. The resulting feature map is then concatenated with the input to the ResNeXt block. Maxpooling with stride (2,2,2) is performed on the ResNeXt output before being passed onto the next layer in the encoding arm. The decoding arm consists of (3×3×3) convolutions with up sampling (stride of [2,2,2]). All the layers use the ReLU activation function except for the output layer, which uses sigmoid.

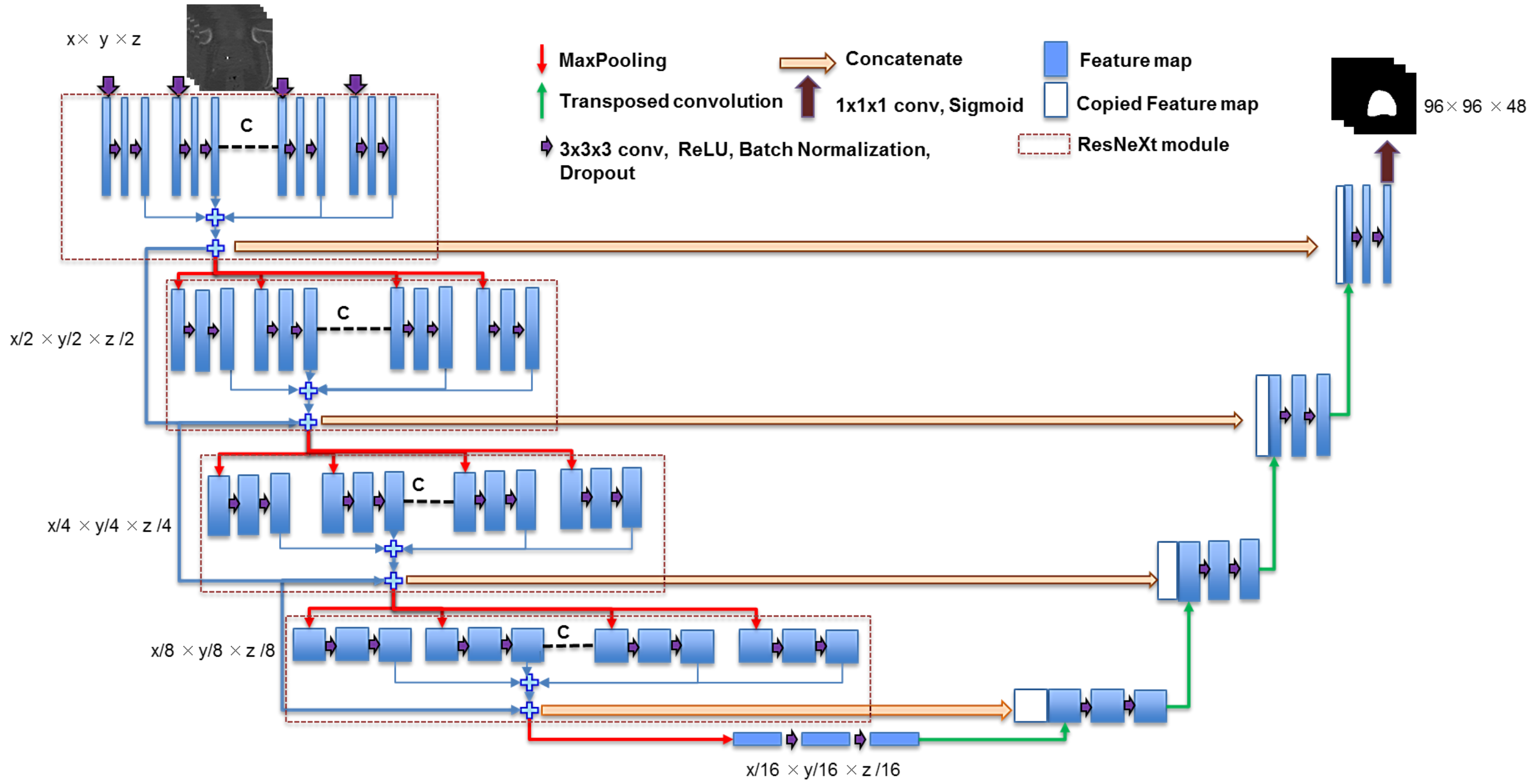

*Figure 3: Segmentation architecture: a 3D U-Net with ResNeXt blocks in the encoding arm.*

The output of the localization network is used for predicting organ volumes that is input into the segmentation network. Since the organ volumes vary from patient to patient and the 3D network requires a common volume size, a fixed number of total slices (32 for prostate and 64 each for the OARs) is selected around the predicted volume. The 3D network was trained by minimizing the boundary weighted dice loss function, $DSC_{bw}$ (described in the next section), using the Adam optimizer [D. P. Kingma et al., 2014]. The number of layers, filter sizes, and also the value of n and hyper parameters varied for different organs.

## Loss functions

The Dice similarity coefficient (*DSC*), also called the overlap index, is the most used metric in validating medical volume segmentations. DSC measures the amount of agreement between two image regions. It is widely used as a metric to evaluate the segmentation performance with the given ground truth in medical images. DSC is defined as

$$DSC = \frac{2|True \cap Pred|}{|True| + |Pred|} \qquad (1)$$

We use | | to indicate the number of foreground voxels in the ground truth (*True*) and predicted (*Pred*) segmentations. DSC is a useful loss function for segmentation networks because of its efficiency in addressing class imbalance issue.

***Organ weighted dice function for localization network***

Since organs vary in size, we have to make sure that all the organs are contributing equally to the dice loss function. To achieve this, a custom weighted Dice function was used for the localization network as defined in (2). All pixels corresponding to annotated areas were assigned a weight inversely proportional to the number of samples of its class in the specific set.

$$DSC_w = \sum\{w_x\} \frac{2|True \cap Pred|}{|True| + |Pred|} \qquad (2)$$

Here $\{w_x\}$ = weights vector for organ x determined during training from the number of samples belonging to the organ. *True* and *Pred* are the gold standard and predicted segmentation results, respectively.

***Boundary weighted dice function for segmentation network***

To mitigate the issue of fuzzy boundary we have used a boundary weighted Dice function defined in (3) for the segmentation network. More importance is given to the pixels near the organ boundary. The closer a pixel is to the boundary, the higher the weight would be and hence larger the loss (1- $DSC_{bw}$).

$$DSC_{bw} = \sum \frac{2|(w_b * \{True \cap Pred\}|}{|\{w_b * True\}| + |\{w_b * Pred\}|} \qquad (3)$$

Here $w_b$ = border weight matrix for the organ determined during training

## EXPERIMENTS & RESULTS

### Patients Data

The dataset consisted of raw CT scan images of 136 prostate cancer patients collected at The University of Texas Southwestern Medical Center (UTSW). All CT images were acquired using a 16-slice CT scanner (Royal Philips Electronics, Eindhoven, The Netherlands). The target organ (prostate) and OARs (bladder, rectum and femoral heads) were contoured by experienced radiation oncologists. All images were acquired with a 512x512 matrix and 2 mm slice thickness (voxel size 1.17mm×1.17mm×2mm).

### Training Data

Eighty percent of the patient data were used for training and the remaining 20% were used for testing. The training data was split into 4 groups and the leave-one-out cross validation strategy was used for training. One set was reserved as the validation data set, and the training was performed to minimize the loss on that validation set. The average of the predictions of the four models was used for predicting the test set.

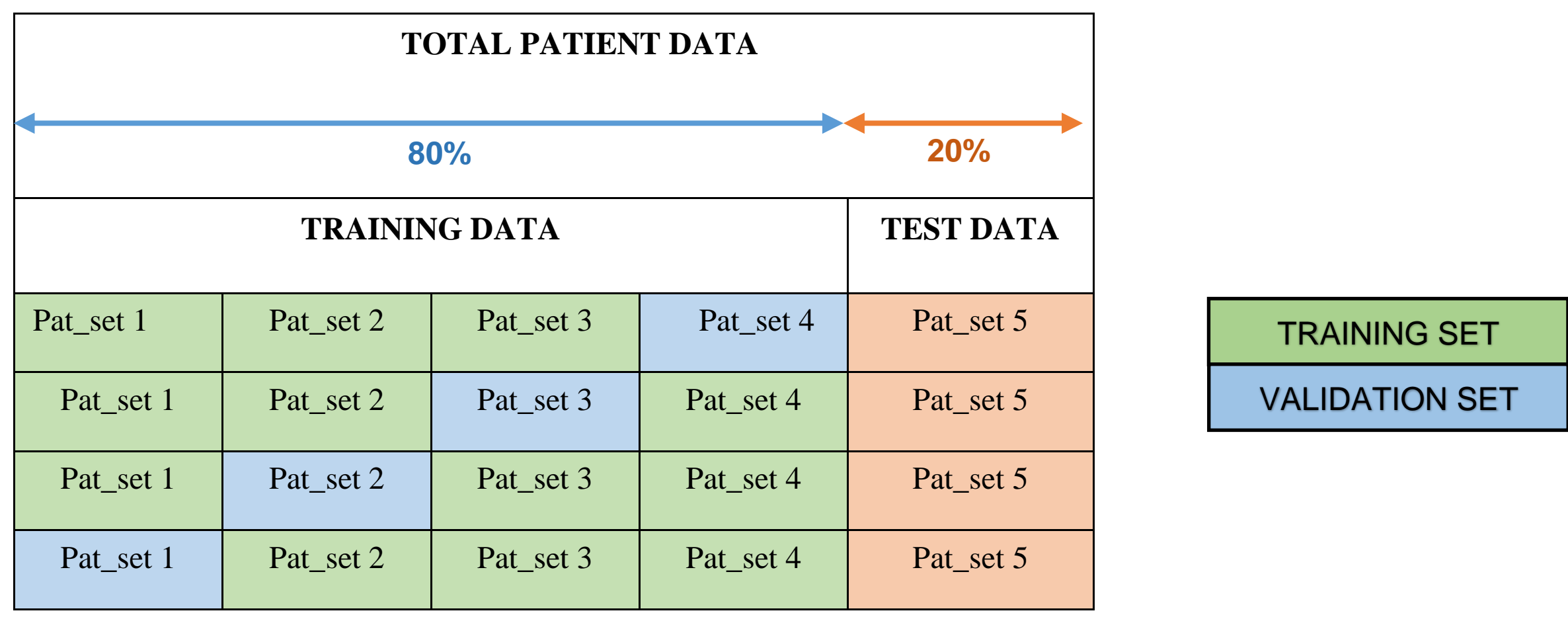

*Figure 4: Leave-one-out cross validation training strategy*

## Training strategies

***Architecture Selection***: In order to design the most optimal network, we applied different CNN architecture variations to UNet. We finalized on a ResNeXt variation of UNet. Using ResNeXT, accuracy can be gained more efficiently by increasing the cardinality than by going deeper or wider. ResNeXt uses a split-transform-merge strategy where inputs are split, then transformed and the outputs of these different paths are merged by adding them together. By applying ResNeXt only along the encoding path of the 3D UNeT, we were able to produce good results without going very deep and keeping computational complexity within what NVIDIA Tesla K80 dual-GPU graphic card can handle for a 3D network. Details of the 2D as well as 3D architectures can be found in Appendix.

***Kernel Size:*** Initially, we tried a kernel size of 5×5×5 for the convolutional layers, but this led the output segmentations to become inaccurate around edges and small contours for the validation data. Intuitively this made sense as we were grouping large sections together and when up-sampling was performed, they did not map to precise points. After trying different kernel sizes of 5×5×5, 3×3×3, and 1×1×1, we settled on 3×3×3 exclusively. The model's accuracy and training speed changed drastically when we used different learning rates. A high learning rate of 1e-2 with a decay of 0.1, when loss plateaus, led to the best validation set Dice coefficient value.

***Overfitting:*** Even though the architecture we employed is efficient enough for accurate prediction, over fitting issues still occur. To overcome over fitting, we used early stopping and dropout [Srivastava, Nitish et al., 2014]. Batch normalization [Sergey Ioffe et al., 2015] was added at each layer to increase network stability. For efficient learning, the learning rate was decayed when validation loss plateaus.

Deep learning models were implemented using the open source Keras package [Chollet, F., 2015], and componential processing was performed with an NVIDIA Tesla K80 dual-GPU graphic card.

## Results

Examples of the segmentations predicted by the model vs the ground truth segmentation are shown in Figures 5 and 6. The segmentations predicted by the automation network (Blue) closely resemble the manually segmentations (Red). Examples of ground truth vs predicted segmentation for base, middle, and apex slices for prostate are illustrated in Figure 7.

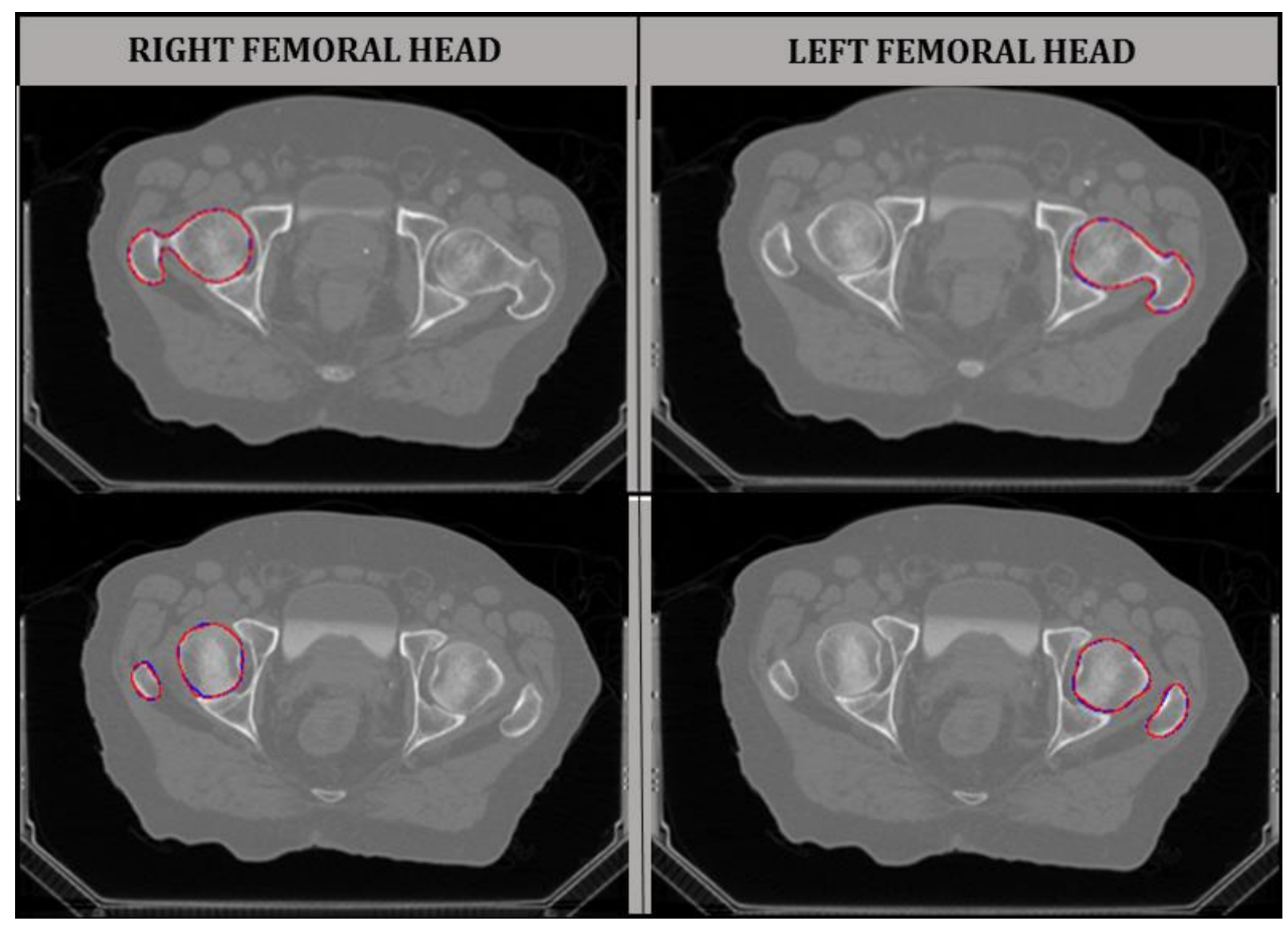

*Figure 5: Predicted segmentation (Blue) vs ground truth (Red) result for left and right femoral head.*

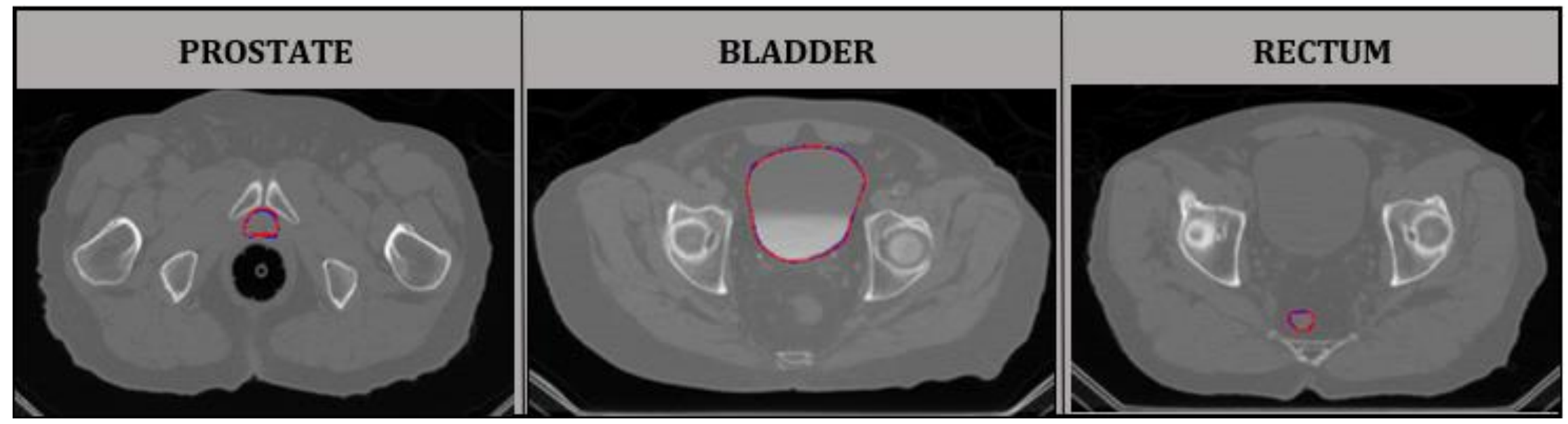

*Figure 6: Predicted segmentation (Blue) vs ground truth (Red) result for prostate, bladder and rectum, from left to right.*

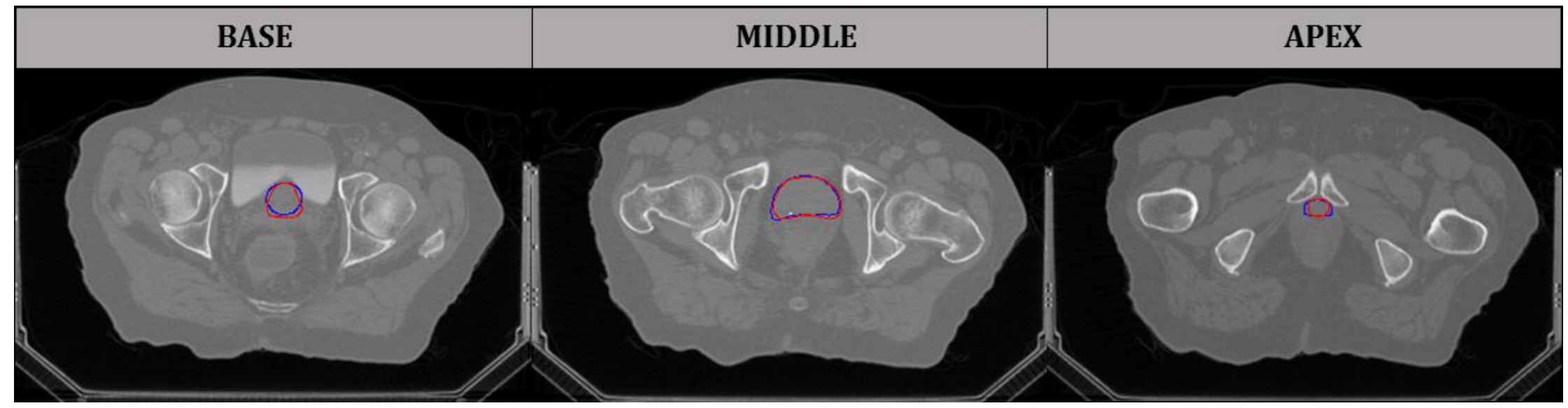

*Figure 7: Predicted segmentation vs. ground truth results for prostate at base, middle, and apex slices, from left to right.*

The calculated DSC values for each of the four cross validation models and the averaged model for prostate segmentation are plotted in Figure 8. With the averaged model, the maximum DSC was 93% and the minimum DSC was 85.9%.

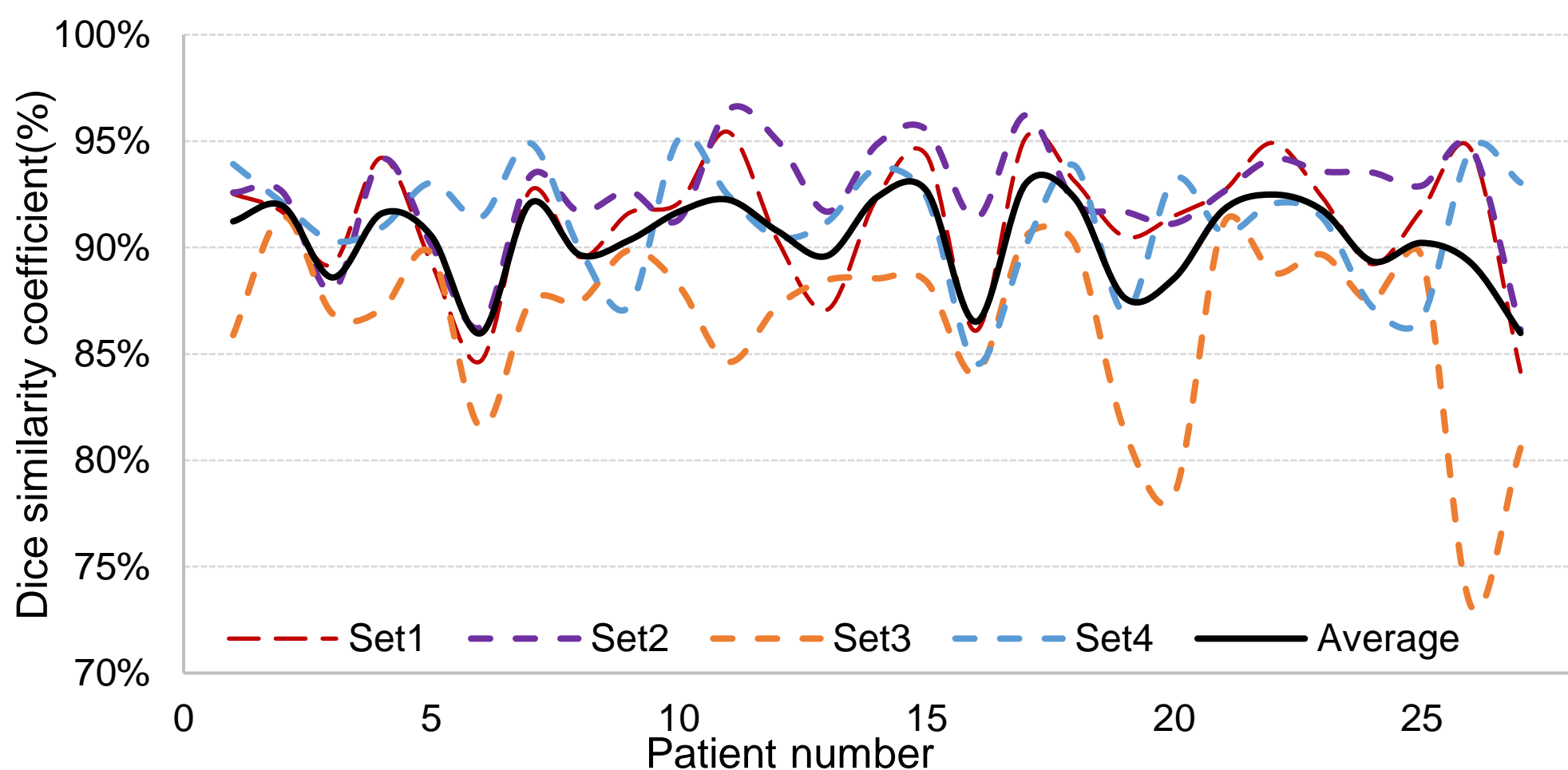

*Figure 8: Plot showing the prostate segmentation DSC values for the cross validation model and the final averaged model.*

The result of cross validation training for bladder, prostate, and rectum is summarized in Table 1. The box whisker plot showing the distribution of Dice coefficient of the organs segmented for all patients is shown in Figure 9.

*Table 1: Cross validation result for bladder, prostate, and rectum.*

| Organ | Accuracy of cross validation Set 1 | Accuracy of cross validation Set 2 | Accuracy of cross validation Set 3 | Accuracy of cross validation Set 4 |
|---|---|---|---|---|
| Bladder | 95.6% | 92.2% | 96.0% | 95.4% |
| Prostate | 91.2% | 90.5% | 88.5% | 91.2% |
| Rectum | 86.5% | 85.4% | 80.2% | 82.4% |

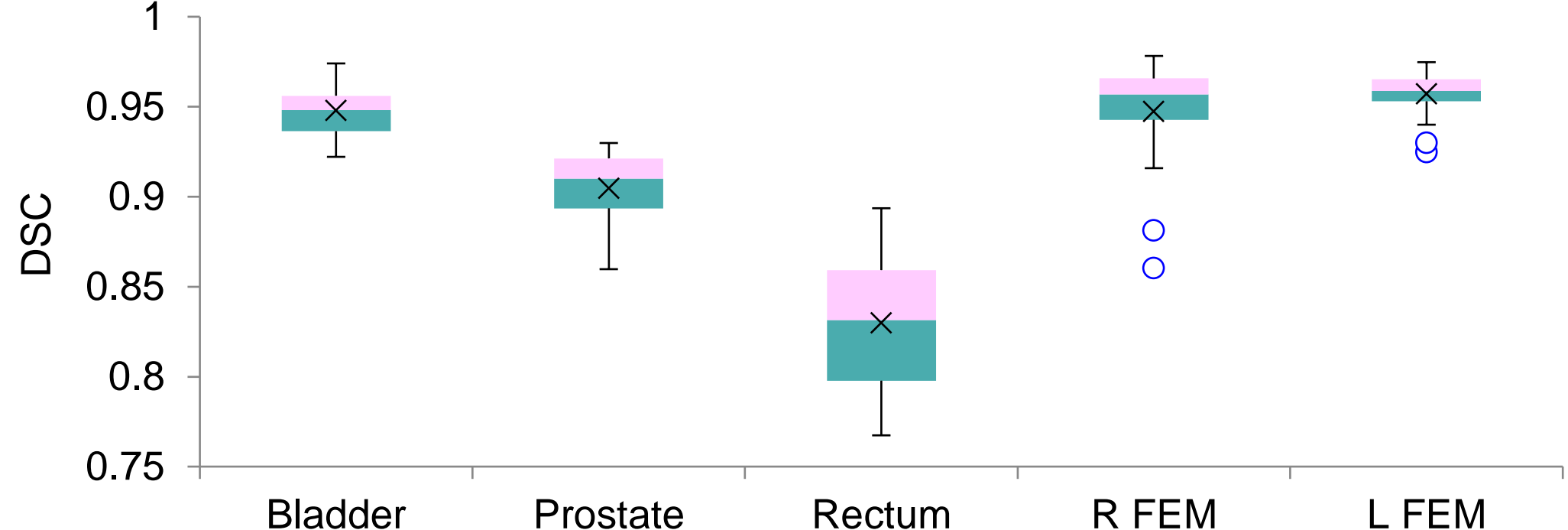

*Figure 9: Box whisker plot showing the distribution of dice coefficient for all the organs segmented.*

The final test accuracy was obtained by averaging the prediction of the four cross validation models on the test dataset. We have also trained the network with a weighted (to compensate for class imbalance) binary cross entropy loss instead of a weighted DSC loss for comparison and have estimated the statistical significance of the improvement in the result when using weighted DSC loss. Average DSC, Average Surface distance and Average Hausdorff distance for the testing data while using DSC loss and, Average DSC while using BCE loss along with the results of significance tests are reported in Table 2.

*Table 2: Mean DSC accuracy for the averaged model and significance test result.*

| | AVERAGE DSC (WITH WEIGHTED DICE LOSS) | AVERAGE DSC (WITH WEIGHTED BINARY CROSS ENTROPY LOSS) | SIGNIFICANCE TEST (WILCOXON SIGNED RANK TEST FOR NON-NORMAL DISTRIBUTION AND PAIRED T-TEST FOR NORMAL DISTRIBUTION) | |
|---|---|---|---|---|
| BLADDER | 95.0(±1.5)% | 90.3(±3.7)% | Z=-3.9748,p=4e-05 | Significant |
| PROSTATE | 90.2(±2.0)% | 84.09(±4.1)% | p = 1.6E-08 | Significant |
| RECTUM | 84.3(±3.7)% | 81.2%(±4.0)% | Z= -1.8098,p=0.035 | Significant |
| RIGHT FEMORAL HEAD | 95.0(±1.3)% | 93.4%(±1.3)% | Z=-1.35, p=0.123 | Not significant |
| LEFT FEMORAL HEAD | 96.0(±3.0)% | 93.8%(±2.9)% | Z=-1.1359, p=0.137 | Not significant |

## DISCUSSION

Accurate segmentation of prostate and OARs is important for radiotherapy treatment planning. The low soft tissue contrast in CT images and organ size, shape, and intensity variations among patients hinder automated segmentation. The proposed fully automated method provides accurate and fast segmentation of prostate and OARs. Once training is completed and the weights for the network are saved, final segmentation for the test set (27 patients) only takes a few seconds. The entire workflow which includes 2D localization, volume cropping, and 3D segmentation only takes a few minutes for each patient. The mean DSC values obtained for prostate, bladder, rectum, left femoral head, and right femoral head were 0.90, 0.95, 0.84, 0.95, and 0.96 respectively. These results indicate high similarity between automated and manual segmentations of the organs. The low mean DSC value as well as the large variation for the rectum is likely due to the large variation in the size and shape of this organ among patients. The dataset included patients with rectal balloon inserts. Even though these cases were predicted well (Figure 10), large variation is introduced into the data. This and the fact that most rectum manual segmentations are performed with the help of MRI may explain the suboptimal results. In the future, we hope that larger datasets will help improve accuracy.

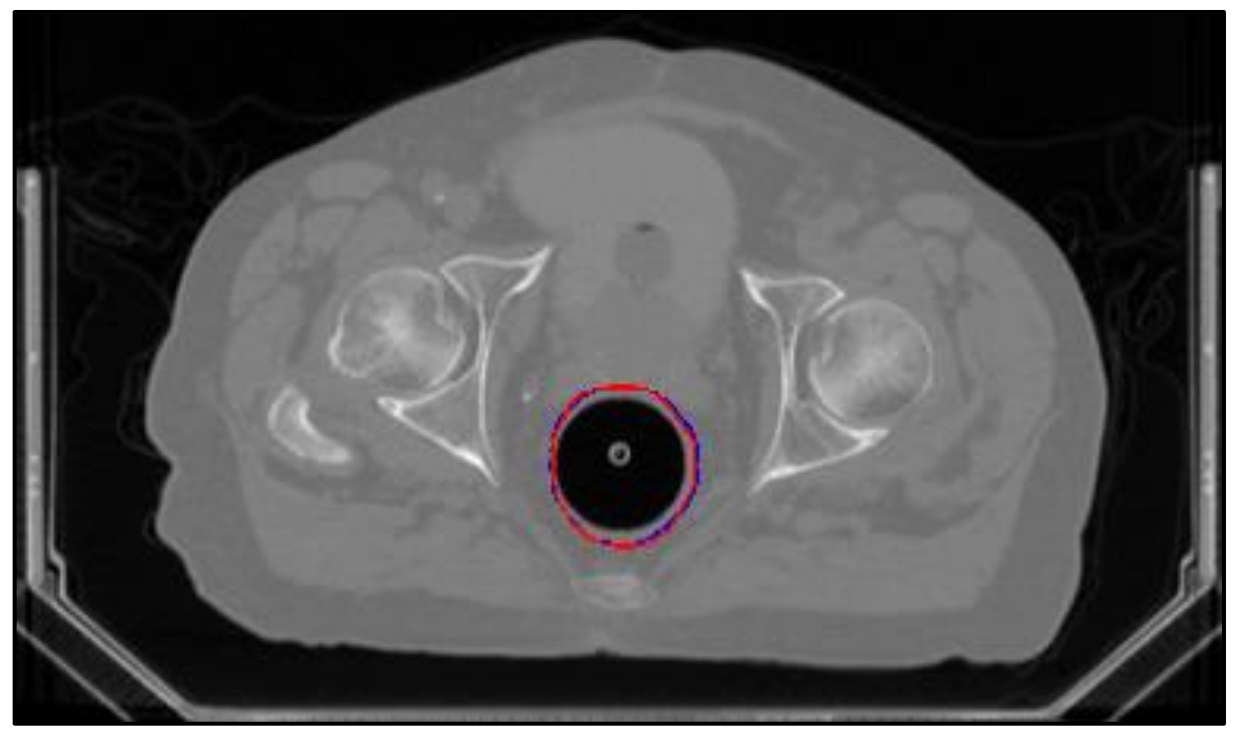

*Figure 10: Predicted segmentation (Blue) vs ground truth (Red) result for a rectum with rectal balloon insert*

Several deformation-based, multi-atlas-based and learning-based methods were developed for the same purpose. A direct comparison is not possible, however, Table 3 shows the comparison between the DSC, Average surface distance (the average distance (in mm) from the ground truth surface to the predicted surface) and Average Hausdorff distance (average of the 90 percentile of surface distances) of our proposed method with the corresponding metrics of the methods that have values reported in their publications.

*Table 3: Comparison of proposed segmentation result with other published results*

| | **Metrics** | **Prostate** | **Bladder** | **Rectum** |
|---|---|---|---|---|
| Shao Y et al. (2014) | **DSC(±SD)%** | 88(±2)% | 86(±8)% | 85(±5)% |
| | **(ASD±SD)mm** | 1.86±0.21 | 2.22±1.01 | 2.21±0.5 |
| | **(HD±SD)mm** | - | - | - |
| Gao Y et al. (2012) | **DSC(±SD)%** | 86(±5)% | 91(±10)% | 79(±20)% |
| | **(ASD±SD)mm** | 1.85 ± 0.74 | 1.71 ± 3.74 | 2.13 ± 2.97 |
| | **(HD±SD)mm** | - | - | - |
| Gao Y et al. (2016) | **DSC(±SD)%** | 87(±4)% | 92(±5)% | 88(±5)% |
| | **(ASD±SD)mm** | 1.77 ± 0.66 | 1.37 ± 0.82 | 1.38 ± 0.75 |
| | **(HD±SD)mm** | - | - | - |
| Feng Q et al. (2010) | **DSC(±SD)%** | 89(±5)% | - | - |
| | **(ASD±SD)mm** | 2.08±0.79 | - | - |
| | **(HD±SD)mm** | - | - | - |
| Martinez et al. (2014) | **DSC(±SD)%** | 87(±7)% | 89(±8)% | 82(±6)% |
| | **(ASD±SD)mm** | - | - | - |
| | **(HD±SD)mm** | 9.98±3.4 | 25.07±4.6 | 13.52 ± 5.1 |
| Ma L et al. (2017) | **DSC(±SD)%** | 86.8(±6.4)% | - | - |
| | **(ASD±SD)mm** | - | - | - |
| | **(HD±SD)mm** | - | - | - |
| Our method | **DSC(±SD)%** | 90(±2.0)% | 95(±1.5)% | 84(±3.7)% |
| | **(ASD±SD)mm** | 0.7±0.5 | 0.5±0.7 | 0.8±0.7 |
| | **(HD±SD)mm** | 5.3±2.8 | 17.0±14.6 | 4.9±3.9 |

Also, we compared the performance of our network with several variations of U-Net model. A comparison of prostate segmentation accuracy and computation time for different segmentation architectures is shown in Table 4. These network configurations were trained and tested using the same patient dataset. Each network was fine-tuned each individually and hence the number of layers, depth and dropout rate values were different for each network architecture. The use of ResNet [Zhang, et al., 2015] improved the results, though they were not significant and involved higher computation times. The performance of DenseNet [Huang, et al., 2018] was even poorer. Our proposed 3D network performed the best, as indicated by a cross validation balanced accuracy of 90.2% for prostate and lowest computation time. The prediction time for all the networks was 6-8s for the test set containing 27 patients. All of these models performed poorly without a boundary weighted Dice loss function (1-$DSC_{bw}$). We present an example of the predicted segmentation result when the network was trained with the boundary weighted dice loss function (1-$DSC_{bw}$) vs when it was trained with normal dice loss function (1-DSC) (Figure 11).

*Table 4: Comparison of the result of different U-Net architectures on the same dataset*

| ***Model*** | ***Computation time (hours for training)*** | ***Prostate segmentation Accuracy(DSC)*** |
|---|---|---|
| *2D U-Net* | 2.0 | 87.0% |
| *3D U-Net* | 2.2 | 89.0% |
| *3D U-Net with 1x1 convolutions* | 2.2 | 88.4% |
| *3D U-Net with DenseNet* | 2.4 | 88.5% |
| *3D U-Net with ResNet* | 2.8 | 89.1% |
| ***3D U-Net with ResNeXt blocks*** | **1.9** | **90.2%** |

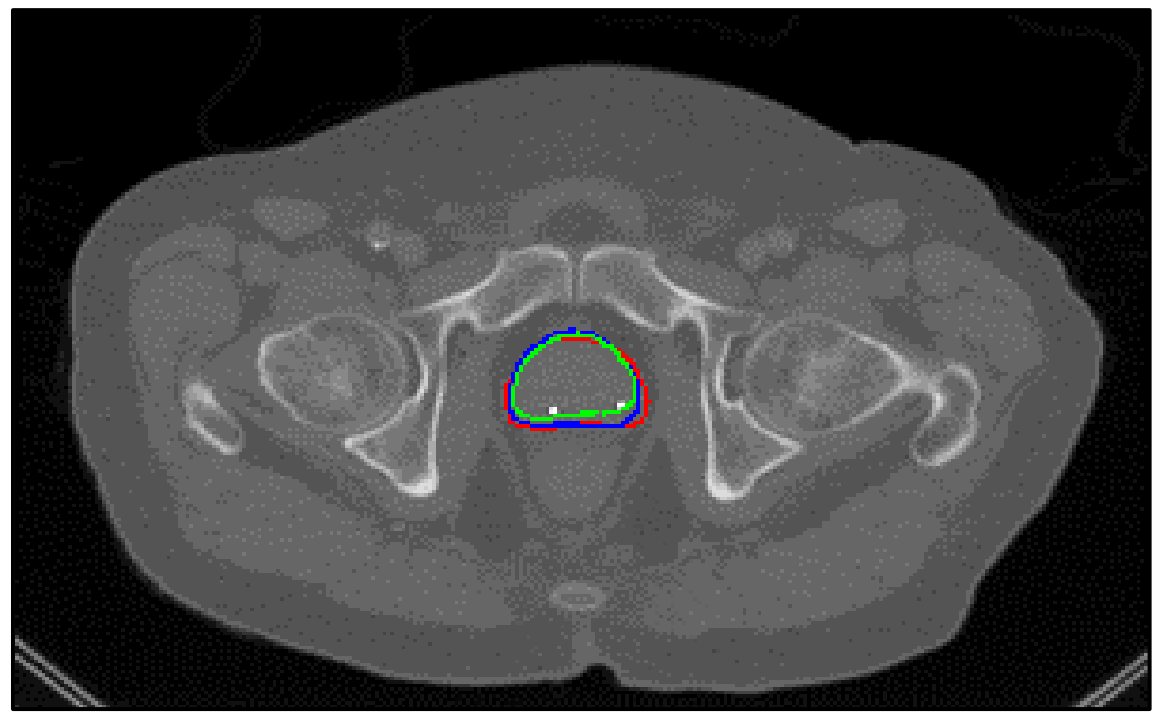

*Figure 11: Predicted segmentation with custom weighted dice loss (blue), with normal dice loss (green), vs ground truth segmentation (red).*

Although it is considered the "gold standard," manual segmentation is not perfect ground truth because of inter-observer variability. In our future work we will ask multiple observers to contour pelvic CT and take an average of the contours as ground truth. Also, we will create an ensemble network to effectively combine the segmentations of all the organs in images without any overlap.

## CONCLUSION

The proposed 2D-3D hybrid network only requires CT images as input, demonstrates a unique ability in dealing with irregular prostates and rectums, and to exhibits superior segmentation performance (i.e., higher DSC value) compared with the state-of-the-art methods (including our own 2D model). Our deep-learning based method is more effective than other automated segmentation methods, and does not require any manual intervention. The proposed workflow could also be easily adapted for segmenting other organs.

## ACKNOWLEDGEMENTS

AB and SJ would like to thank the Cancer Prevention and Research Institute of Texas (CPRIT) for their financial support through grant IIRA RP150485. The authors thank Dr. Damiana Chiavolini for proof-reading and editing the manuscript.

## Appendix A. Details on the 5 channel 2D U-Net architectures used for organ localization

| | 5 channel 2D U-Net architecture | |
|---|---|---|
| Layer number | Layer type | Number of features |
| 1 | Input | 5 |
| 2 | Conv | 8 |
| 3 | Conv | 8 |

| 4 | Conv | 8 |
|---|---|---|
| 5 | Max Pooling | 8 |
| 6 | Conv | 16 |
| 7 | Conv | 16 |
| 8 | Conv | 16 |
| 9 | Max Pooling | 16 |
| 10 | Conv | 32 |
| 11 | Conv | 32 |
| 12 | Conv | 32 |
| 13 | Max Pooling | 32 |
| 14 | Conv | 64 |
| 15 | Conv | 64 |
| 16 | Conv | 64 |
| 17 | Conv | 128 |
| 18 | Conv | 128 |
| 19 | U-Net Upsample | 128 |
| 20 | Conv | 64 |
| 21 | Conv | 64 |
| 22 | U-Net Upsample | 64 |
| 23 | Conv | 32 |
| 24 | Conv | 32 |
| 25 | U-Net Upsample | 32 |
| 26 | Conv | 16 |
| 27 | Conv | 16 |
| 28 | U-Net Upsample | 16 |
| 29 | Conv | 8 |
| 30 | Conv | 8 |
| 31 | Conv | 5 |

## Appendix B. Details on the modified 3D U-Net architectures used for organ segmentation

| Modified 3D U-Net architecture | | | | | | | | |
|---|---|---|---|---|---|---|---|---|
| | PROSTATE (n=4) | | BLADDER (n=2) | | RECTUM (n=4) | | FEMORAL HEADS (n=2) | |
| EPOCHS | ~350 | | ~250 | | ~400 | | ~170 | |
| Layer number | Layer type | Number of features | Layer type | Number of features | Layer type | Number of features | Layer type | Number of features |
| 1 | Input | 1 | Input | 1 | Input | 1 | Input | 1 |

| 2 | Conv | 8 | Conv | 4 | Conv | 8 | Conv | 4 |
|---|---|---|---|---|---|---|---|---|
| 3 | Conv | 2 | Conv | 2 | Conv | 2 | Conv | 2 |
| 4 | Conv | 2 | Conv | 2 | Conv | 2 | Conv | 2 |
| 5 | Conv | 2 | Concatenate | 4 | Conv | 2 | Concatenate | 4 |
| 6 | Conv | 2 | Add | 4 | Conv | 2 | Add | 4 |
| 7 | Concatenate | 8 | Max Pooling | 4 | Concatenate | 8 | Max Pooling | 4 |
| 8 | Add | 8 | Conv | 8 | Add | 8 | Conv | 8 |
| 9 | Max Pooling | 8 | Conv | 4 | Max Pooling | 8 | Conv | 4 |
| 10 | Conv | 16 | Conv | 4 | Conv | 16 | Conv | 4 |
| 11 | Conv | 4 | Concatenate | 8 | Conv | 4 | Concatenate | 8 |
| 12 | Conv | 4 | Add | 8 | Conv | 4 | Add | 8 |
| 13 | Conv | 4 | Max Pooling | 8 | Conv | 4 | Max Pooling | 8 |
| 14 | Conv | 4 | Conv | 16 | Conv | 4 | Conv | 16 |
| 15 | Concatenate | 16 | Conv | 8 | Concatenate | 16 | Conv | 8 |
| 16 | Add | 16 | Conv | 8 | Add | 16 | Conv | 8 |
| 17 | Max Pooling | 16 | Concatenate | 16 | Max Pooling | 16 | Concatenate | 16 |
| 18 | Conv | 32 | Add | 16 | Conv | 32 | Add | 16 |
| 19 | Conv | 8 | Max Pooling | 16 | Conv | 8 | Max Pooling | 16 |
| 20 | Conv | 8 | Conv | 32 | Conv | 8 | Conv | 32 |
| 21 | Conv | 8 | Conv | 16 | Conv | 8 | Conv | 16 |
| 22 | Conv | 8 | Conv | 16 | Conv | 8 | Conv | 16 |
| 23 | Concatenate | 32 | Concatenate | 32 | Concatenate | 32 | Concatenate | 32 |
| 24 | Add | 32 | Add | 32 | Add | 32 | Add | 32 |
| 25 | Max Pooling | 32 | Max Pooling | 32 | Max Pooling | 32 | Max Pooling | 32 |
| 26 | Conv | 64 | Conv | 64 | Conv | 64 | Conv | 64 |
| 27 | Conv | 16 | Conv | 32 | Conv | 16 | Conv | 32 |
| 28 | Conv | 16 | Conv | 32 | Conv | 16 | Conv | 32 |
| 29 | Conv | 16 | Concatenate | 64 | Conv | 16 | Concatenate | 64 |
| 30 | Conv | 16 | Add | 64 | Conv | 16 | Add | 64 |
| 31 | Concatenate | 64 | Conv | 128 | Concatenate | 64 | Conv | 128 |
| 32 | Add | 64 | Conv | 64 | Add | 64 | Conv | 64 |
| 33 | Max Pooling | 64 | Conv | 64 | Max Pooling | 64 | Conv | 64 |
| 34 | Conv | 128 | Concatenate | 128 | Conv | 128 | Concatenate | 128 |

| 35 | Conv | 64 | Add | 128 | Conv | 64 | Add | 128 |
|---|---|---|---|---|---|---|---|---|
| 36 | Conv | 64 | 3D Upsample | 128 | Conv | 64 | 3D Upsample | 128 |
| 37 | Conv | 64 | Conv | 64 | Conv | 64 | Conv | 64 |
| 38 | Conv | 64 | Conv | 64 | Conv | 64 | Conv | 64 |
| 39 | Concatenate | 128 | 3D Upsample | 64 | Concatenate | 128 | 3D Upsample | 64 |
| 40 | Add | 128 | Conv | 32 | Add | 128 | Conv | 32 |
| 41 | 3D Upsample | 128 | Conv | 32 | 3D Upsample | 128 | Conv | 32 |
| 42 | Conv | 64 | 3D Upsample | 32 | Conv | 64 | 3D Upsample | 32 |
| 43 | Conv | 64 | Conv | 16 | Conv | 64 | Conv | 16 |
| 44 | 3D Upsample | 64 | Conv | 16 | 3D Upsample | 64 | Conv | 16 |
| 45 | Conv | 32 | 3D Upsample | 16 | Conv | 32 | 3D Upsample | 16 |
| 46 | Conv | 32 | Conv | 8 | Conv | 32 | Conv | 8 |
| 47 | 3D Upsample | 32 | Conv | 8 | 3D Upsample | 32 | Conv | 8 |
| 48 | Conv | 16 | 3D Upsample | 8 | Conv | 16 | 3D Upsample | 8 |
| 49 | Conv | 16 | Conv | 4 | Conv | 16 | Conv | 4 |
| 50 | 3D Upsample | 16 | Conv | 4 | 3D Upsample | 16 | Conv | 4 |
| 51 | Conv | 8 | Conv | 1 | Conv | 8 | Conv | 1 |
| 52 | Conv | 8 | | | Conv | 8 | | |
| 53 | Conv | 1 | | | Conv | 1 | | |

## Appendix C. Learning curve for a single validation set for prostate segmentation

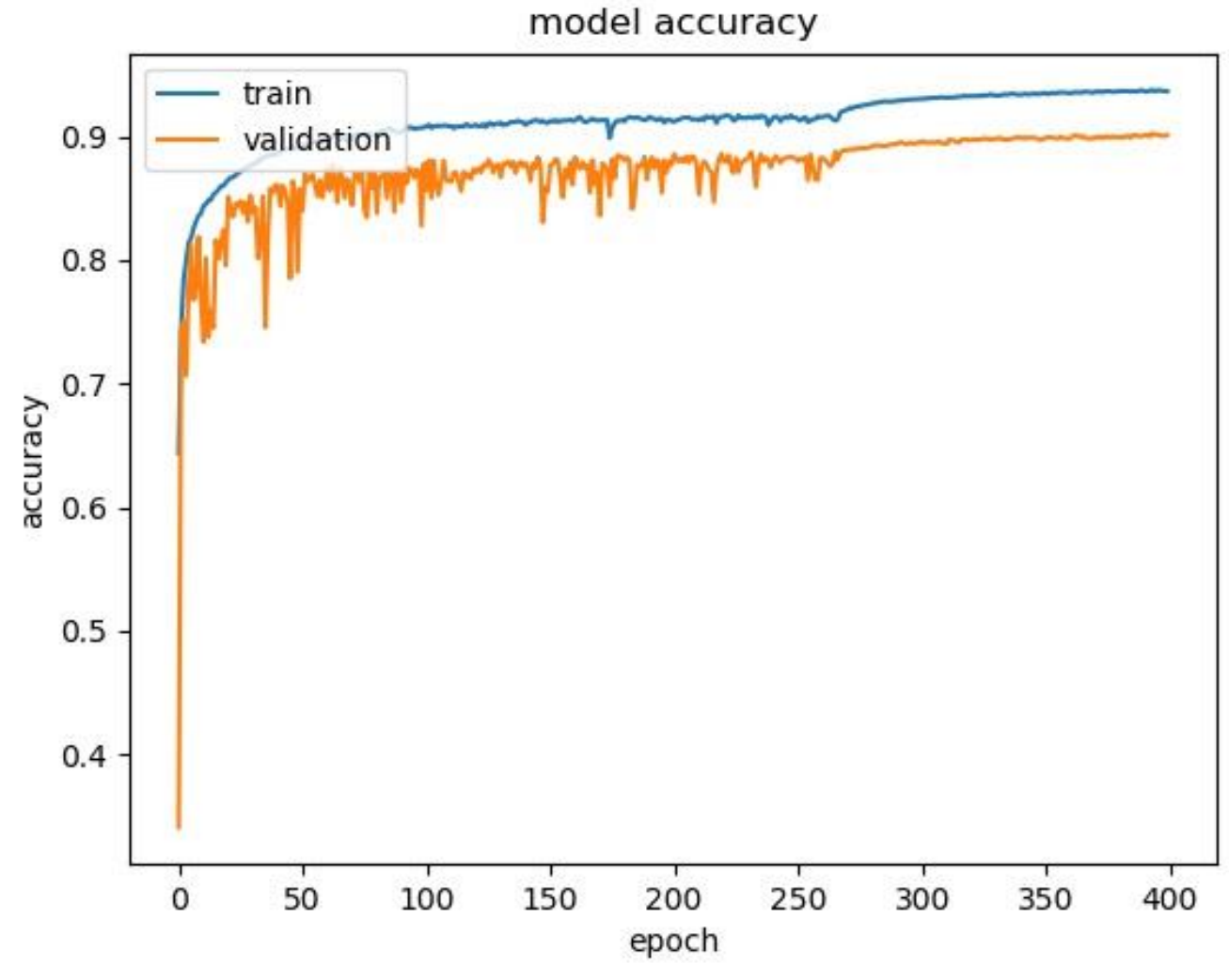

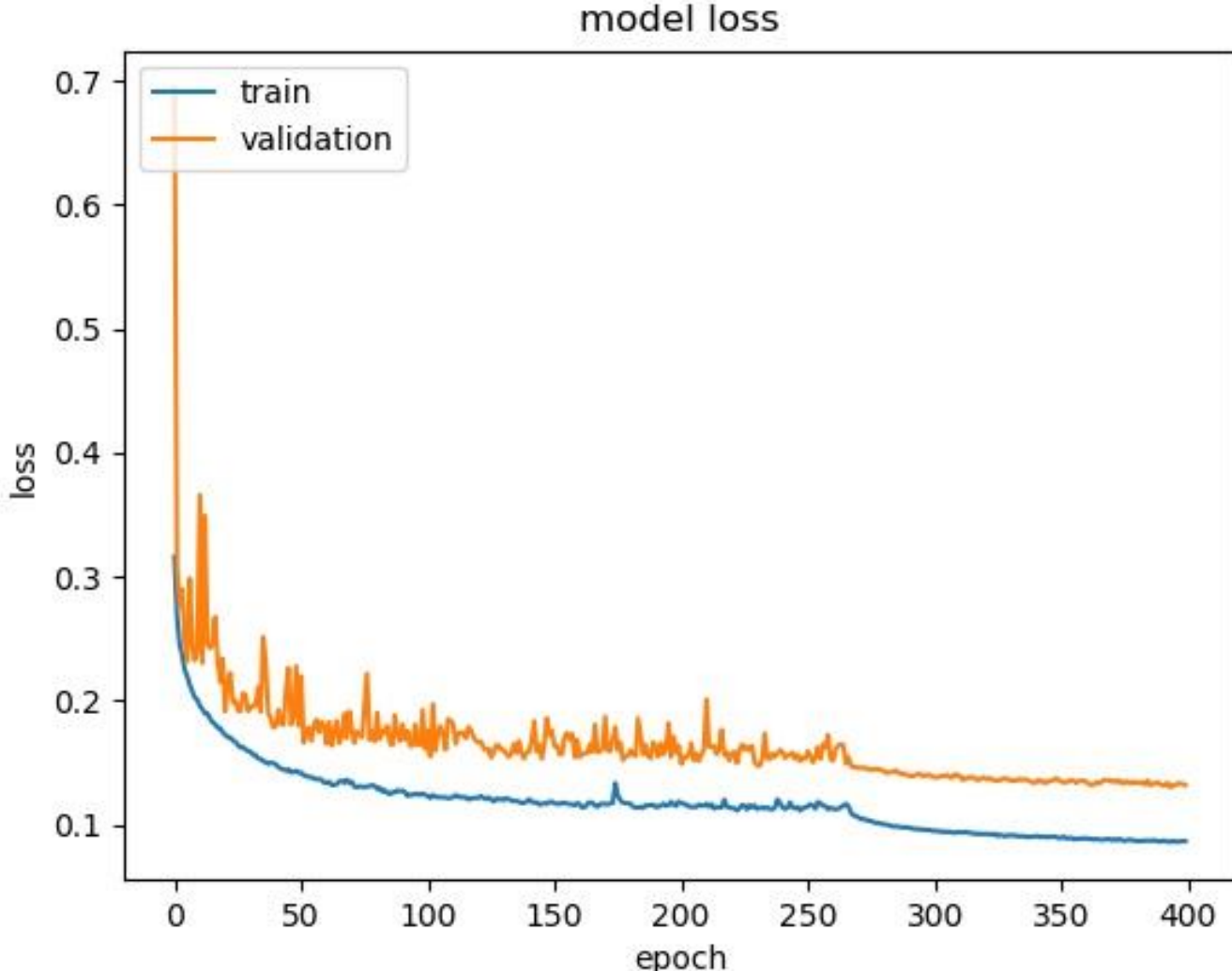